\documentclass{article}

\usepackage{arxiv}

\usepackage[utf8]{inputenc} 
\usepackage[T1]{fontenc}    
\usepackage{hyperref}       
\usepackage{url}            
\usepackage{booktabs}       
\usepackage{amsfonts}       
\usepackage{nicefrac}       
\usepackage{microtype}      
\usepackage{lipsum}
\usepackage{graphicx}
\usepackage{amsmath}
\usepackage{array}
\usepackage{multirow}
\usepackage{colortbl}
\usepackage{xcolor}
\usepackage{pifont}
\usepackage{epstopdf}
\graphicspath{ {./images/} }

\title{MorphoGuard: A Morphology-Based Whole-Body Interactive Motion Controller}

\author{
 Chengjin Wang \\
  Shanghai Research Institute for Intelligent Autonomous Systems\\
  Shanghai 201210, China\\
  \texttt{2210991@tongji.edu.cn} \\
   \And
 Zheng Yan \\
  Shanghai Research Institute for Intelligent Autonomous Systems\\
  Shanghai 201210, China\\
  \texttt{2310292@tongji.edu.cn} \\
  \And
Yanmin Zhou \\
  College of Electronics and Information Engineering (Tongji University)\\
  Shanghai 201210, China\\
  \texttt{yanmin.zhou@tongji.edu.cn} \\
  \And
Runjie Shen \\
  College of Electronics and Information Engineering (Tongji University)\\
  Shanghai 201210, China\\
  \texttt{11132@tongji.edu.cn} \\
  \And
Bin He \\
  College of Electronics and Information Engineering (Tongji University)\\
  Shanghai 201210, China\\
  \texttt{hebinhe@tongji.edu.cn} \\
}

\begin{document}
\maketitle

\begin{figure}[htbp]
\centering
\includegraphics[width= \textwidth]{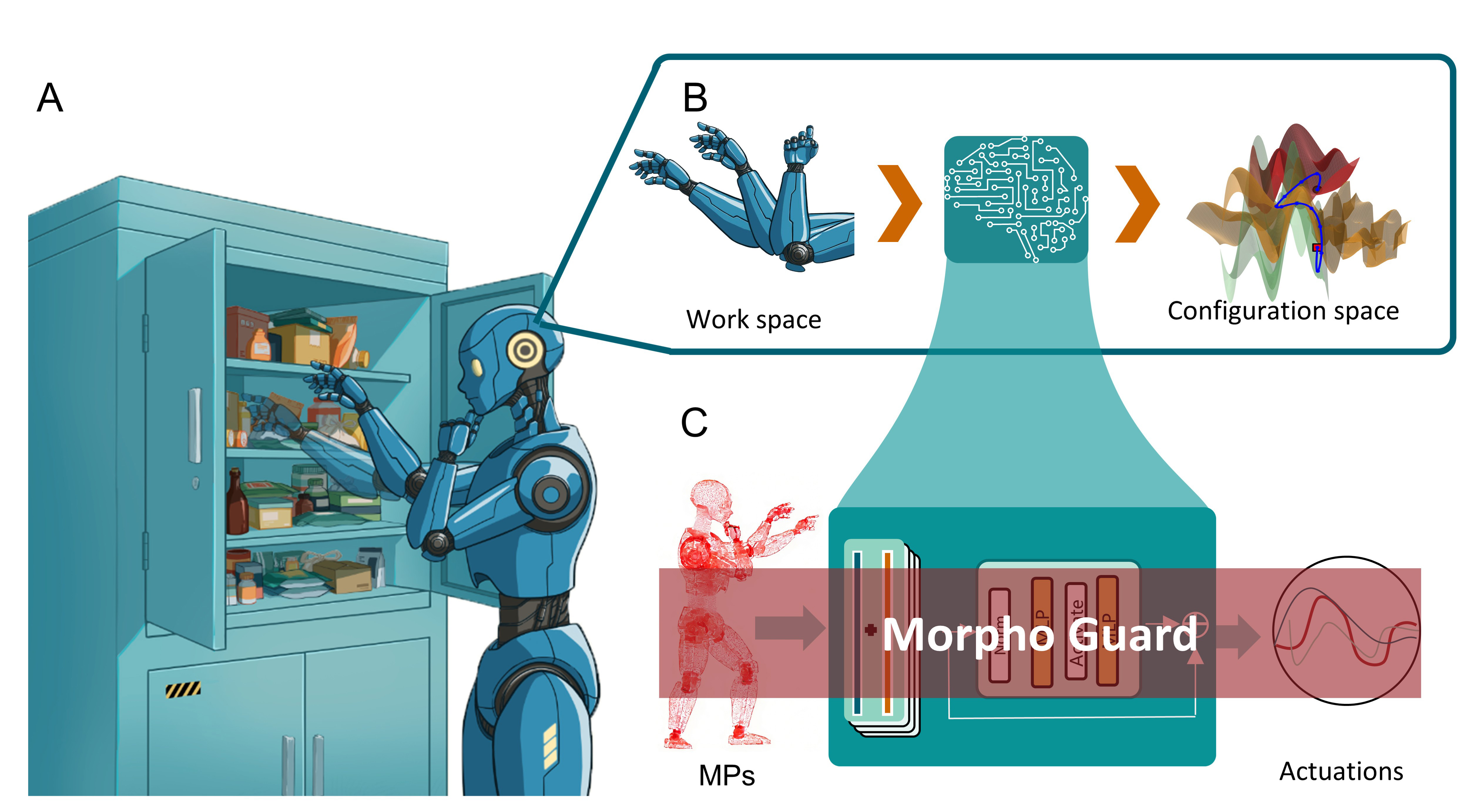} 
\caption{Schematic of morphology-based whole-body motion control (MorphoGuard). (A) An example of a robot facing complex contact combinations in an unstructured environment. (B) MorphoGuard is tasked with learning a mapping from the entity morphology in the workspace to the configuration space. (C) MorphoGuard takes as input MPs that discretely represent the robot morphology and predicts the corresponding joint motion commands.}
\label{fig:1}

\end{figure}
\begin{abstract}
Whole-body control (WBC) has demonstrated significant advantages in complex interactive movements of high-dimensional robotic systems. However, when a robot is required to handle dynamic multi-contact combinations along a single kinematic chain—such as pushing open a door with its elbow while grasping an object—it faces major obstacles in terms of complex contact representation and joint configuration coupling. To address this, we propose a new control approach that explicitly manages arbitrary contact combinations, aiming to endow robots with whole-body interactive capabilities. We develop a morphology-constrained WBC network (MorphoGuard)—which is trained on a self-constructed dual-arm physical and simulation platform. A series of model recommendation experiments are designed to systematically investigate the impact of backbone architecture, fusion strategy, and model scale on network performance. To evaluate the control performance, we adopt a multi-object interaction task as the benchmark, requiring the model to simultaneously manipulate multiple target objects to specified positions. Experimental results show that the proposed method achieves a contact point management error of approximately 1 cm, demonstrating its effectiveness in whole-body interactive control.
\end{abstract}


\section{Introduction}
	
    Whole-Body Control (WBC) has demonstrated significant potential in enhancing the motion coordination and safety of high-dimensional robotic systems, particularly during complex physical interactions \cite{khatib2022constraint, sentis2006whole, barreiros2025learning,simetti2015whole, fu2023deep, li2024realizing, jiang2024learning}. Existing motion strategies primarily focus on optimizing interactive motions at the endpoints of multiple kinematic chains, for example, the coordinated movement between a robotic arm's end-effector and the feet \cite{li2024realizing, jiang2024learning}. However, research on complex contact combinations along a single kinematic chain remains relatively limited. In daily activities, it is common to encounter tasks that require coordinated efforts from multiple parts of a limb to accomplish complex interactions, such as collaboratively carrying a large box using both the upper arm and forearm, or using an elbow to push open a door while holding an object. The core challenge of such contact combinations stems from the coupling of joint configurations, which directly leads to stiffness and discontinuities in the search space. Consequently, previous studies have often addressed these issues by incorporating them into skill-level motion policies through learning from expert demonstrations.

	In this paper, we propose a morphology-guaranteed WBC method to explicitly manage such complex contact combinations. The advantage of morphological representation lies in its ability to uniformly represent multiple interaction interfaces in a geometric form \cite{chen2022fully,sun2023embedded,belke2023morphological,diaz2023machine}. This homogenized representation effectively avoids the coupling of joint configurations caused by multi-contact interactions. Leveraging the homeomorphic mapping between the workspace (i.e., the set of achievable morphological states of the robot) and the configuration space (Fig. \ref{fig:1}B), we construct a unique inverse kinematics mapping. This embodied model, oriented toward the physical entity, endows the robot with morphological adaptation capabilities, enabling it to explicitly manage contact combinations. The key to achieving such morphological control lies in ensuring a spatiotemporally consistent representation of the robot's topological features. As the robot's morphology deforms with changes in joint configurations, it is necessary to register interactive perception signals with the time-varying morphology. More critically, the inverse kinematics modeling of the robot relies on spatially topological features that maintain temporal consistency.

	To address this challenge, we extend the material point method to the morphological representation of robots, as shown in Fig. \ref{fig:1}C \cite{de2020material, fuchslin2013morphological}. This method has achieved notable success in simulating the motion and deformation of various objects. Through spatial discretization \cite{skeel1990method,braun1997modelling}, the robot morphology is represented as a finite set of material points (MPs) with fixed topological relationships. These MPs, which carry physical quantities, lay the foundation for modeling the complex motion characteristics of the robot. Specifically, we bind the electronic skin\cite{guo2021visual, zhong2024high}—characterized by a fixed spatial distribution—to the MPs to capture interaction stimuli. External stimuli, together with intrinsic actuation, constitute the boundary conditions for the MPs. By tracking the interface of the motion states of these MPs, the robot's morphology can be represented explicitly with fine granularity.

\begin{figure}[htbp]
	\centering
	\includegraphics[width= 0.8 \textwidth]{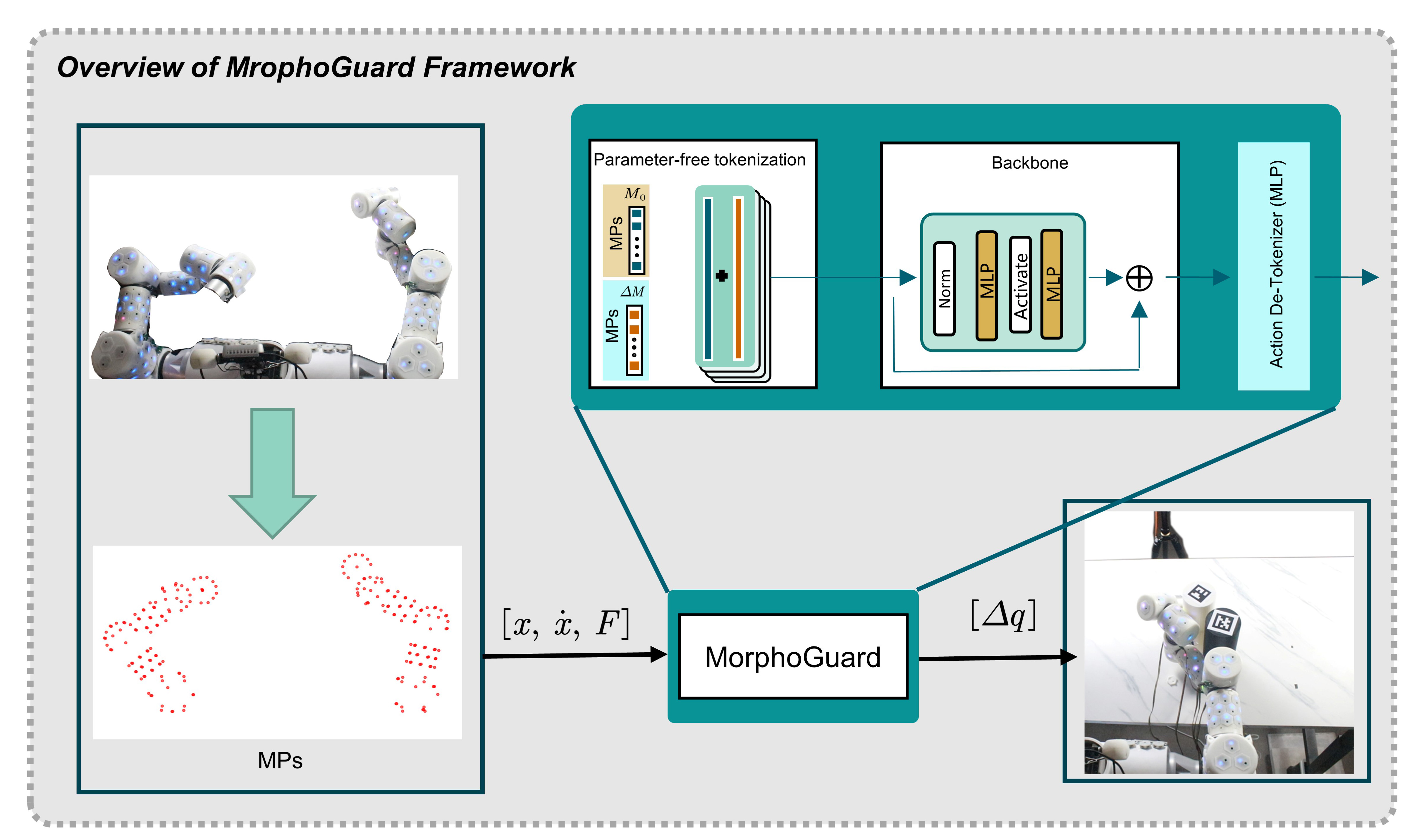} 
	\caption{ An overview of the MorphoGuard framework. The model consists of a MPs encoding module, an fusion module, and a joint command decoding module. The morphology encoding module encodes the current morphology and the target morphology into feature representations. The information fusion module integrates the encoded features to capture the relationship between the current and target morphologies. Finally, the joint command decoding module predicts the joint commands based on the fused features.}
	\label{fig:2}
	\end{figure}

	Based on this, we propose a deep neural network model termed MorphoGuard. The model adopts an encoder-decoder architecture designed to encode the spatial states of the robot morphology and predict the corresponding joint commands as shown in Fig. \ref{fig:1}. To achieve a spatiotemporally consistent representation of the morphological geometry, we introduce a MPs spatial state feature auto-encoding module. Built upon this backbone network, the decoder is responsible for predicting the joint commands that correspond to the given geometric configuration. The overall architecture of the model is illustrated in Fig. \ref{fig:2}.

	To achieve this goal, we first established a data acquisition platform for model training. The platform comprises a dual-arm robot fully covered with electronic skin, which can operate in both real-world and simulation environments \cite{zhou2023tacsuit}. By controlling the robot to track given joint configurations, we constructed morphology-boundary conditions data pairs under various motion states. Through exhaustive traversal of the robot's workspace, a total of 1.3 M training samples were collected for subsequent model training. We further evaluated the robustness of the proposed model on a multi-body manipulation task. Experimental results show that the model achieves a joint control error of 0.5 and a contact point management error of 1 cm.

\section{Related Works}

	\textbf{WBC} aims to coordinate the behaviors of high-dimensional robotic systems under complex motion constraints, thereby overcoming the motion coordination challenges inherent in traditional decoupled motion modeling approaches \cite{khatib2022constraint, sentis2006whole, simetti2015whole,barreiros2025learning, fu2023deep,li2024realizing,jiang2024learning}. Early research primarily focused on model-based methods, which manage the coupling relationships among multiple kinematic chains through precise dynamic modeling and optimization frameworks \cite{khatib2022constraint, sentis2006whole}. However, such approaches often rely on strong modeling assumptions, limiting their robustness in unstructured environments.

	To address these limitations, learning-based methods have been progressively introduced into the field of WBC over the past two years \cite{barreiros2025learning, alirezazadeh2022dynamic, fu2023deep}. By training motion policies, researchers have enabled robots to exhibit impressive complex behaviors in real-world settings. Nevertheless, existing learning-based motion policies still primarily focus on motion optimization at the endpoints of multiple kinematic chains. This end-effector-oriented design results in a lack of explicit whole-body contact management capability within the policy model. When a robot needs to interact with the environment using other body parts—such as the upper arm, elbow, or torso—the input space of the policy network struggles to effectively represent and utilize such contact information, thereby limiting the robot's performance in tasks involving complex contact combinations \cite{barreiros2025learning}. A potential solution lies in incorporating morphological information directly into the input space of the motion control policy \cite{chen2022fully, sun2023embedded}. This morphological representation can explicitly characterize contact states at arbitrary body positions, thereby endowing the control policy with the capability to manage contacts between any point on the robot body.

	\textbf{Morphological Computation.} This method refers to the process by which a robotic system reasons about its own shape and motion characteristics \cite{xiong2023universal, sun2023embedded, chen2022fully, sun2023embedded,belke2023morphological, fuchslin2013morphological}. Morphological computation has shown significant potential across multiple domains, such as predicting collisions \cite{ belke2023morphological, chen2022fully, sun2023embedded} and facilitating the transfer of control policies between heterogeneous robots \cite{xiong2023universal, lyu2024odrl,li2024mat}. Meanwhile, other works concentrate on directly controlling the compliant actuation structures of robots, driving body motion by modulating morphological changes \cite{fuchslin2013morphological, yuan2026iksel}.

	In contrast, our focus is on morphology-based joint command prediction. Building upon this, the robot can manage interactions at arbitrary body positions.

	\textbf{Learning inverse kinematics.} This method try to derive mappings from task space to joint space directly from data using statistical learning algorithms \cite{aristidou2018inverse,d2001learning,ren2020learning,zhao2023learning,ortone2026bioinspired}. This approach exhibits distinct advantages in processing complex or ambiguous input information. Building on this paradigm, researchers address challenges posed by kinematic redundancies through augmenting input information \cite{d2001learning, ren2020learning, lauretti2022geometric}. Furthermore, enforcing temporally dependent representations enables models to capture dynamic characteristics. These techniques also enhance adaptability to hardware variability and low-precision issues commonly encountered in real-world deployment, such as joint wear and morphological deformation \cite{zhao2023learning, ortone2026bioinspired}.

	We introduce this method into this work. By leveraging the rich morphological information within a neural network architecture, we aim to achieve a robust and efficient inverse kinematics solving model.


\section{Preliminary and Proposal Statements}
\label{sec:pre_state}

	\textbf{WBC}. Consider a robotic system $X(t)$ with $n$ degrees of freedom. The objective of WBC is to propose a control policy $\mathcal{K}$ based on a unified motion model. Given the current motion state $x_t$, this policy computes the actuation $u \in T^n$ required to track the desired spatial state $x_{t+1}$ at the next time step. In this work, the motion state is extended from the end-effector trajectory $x: p \in \mathbb{R}^3 $ to the robot's morphological state $x: \mathcal{M} \subset \mathbb{R}^3$, enabling a homogenized representation of contact combinations. Accordingly, the morphology-based WBC task is formulated as $u = \mathcal{K}(\mathcal{M}_t, \mathcal{M}_{t+1})$.

	In this work, to make morphological representation practically feasible, we discretize the robot body into a set of $ M $ MPs, thereby constructing a geometric configuration space $ \mathcal{M} $, with the corresponding joint configuration space denoted as $ \mathcal{Q} $. When the robot is in a joint configuration state $ \mathbf{q} \in \mathcal{Q} $, its corresponding motion morphology is given by $ \mathbf{M}_0 \in \mathcal{M} $. We define the state mapping from the current morphology to the target morphology as $ \mathcal{F}: \mathcal{Q} \times \mathcal{M} \rightarrow \mathcal{M} $, where $ \mathcal{F} $ is clearly twice differentiable. Given the initial coordinates $ \mathbf{M}_0 $ and the target coordinates $ \mathbf{M}_g $ of the material point set, the WBC problem based on discrete morphological representation can be formulated as solving for the joint configuration state $ \mathbf{q} $ such that $ \mathcal{F}(\mathbf{q}) = \mathbf{M}_g $. For ease of presentation, we denote this motion control solving equation as $ \Gamma( \mathbf{M}_0, \mathbf{M}_g) \rightarrow \mathbf{q}$.

	\textbf{Learning-Based Morphology Motion Modeling.} The goal of this problem is to leverage the powerful learning capabilities of neural networks to learn a set of parameters $\mathbf{\theta}$ that describe the morphological motion characteristics of robots in complex, high-dimensional spaces. Leveraging this parametric model, the robot can predict the joint motion commands $\mathbf{q}$ that enable the transition from an initial morphological state $\mathbf{M}_0$ to a target morphological state $\mathbf{M}_g$ within a control error tolerance $e$. Specifically, the task of the network is to learn the conditional probability distribution $p_{\Theta}(\mathbf{q} | \mathbf{z}_0, \mathbf{z}_g)$ of the joint motion commands $\mathbf{q}$, given the current morphological features $\mathbf{z}_0$ and the target morphological features $\mathbf{z}_g$. Formally, this problem can be expressed as the conditional probability distribution:
	\begin{equation}
		p_{\Theta}(\mathbf{q} | \mathbf{M}_0, \mathbf{M}_g) = \mathcal{N}(\mathbf{q}; \mu_{\Theta}(\mathbf{z}_0, \mathbf{z}_g), \Sigma_{\Theta}(\mathbf{z}_0, \mathbf{z}_g)),
	\end{equation}
	where $\mathbf{z}_0=\Phi(\mathbf{M}_0) + \epsilon$ and $\mathbf{z}_g=\Phi(\mathbf{M}_g) + \epsilon$ $(\epsilon \sim \mathcal{N}(0, \sigma^2))$, $\mu_{\Theta}$ and $\Sigma_{\Theta}$ are the model parameters.

	\textbf{Uniqueness of Morphological Parameter Solutions.} Since $ f: \mathcal{Q} \to \mathcal{M} $ is a homeomorphism, its inverse mapping $ f^{-1}: \mathcal{M} \to \mathcal{Q} $ exists and is continuous. Given $ \mathbf{M}_0, \mathbf{M}_g \in \mathcal{M} $, let $ \mathbf{q}_0 = f^{-1}(\mathbf{M}_0) $ and $ \mathbf{q}_g = f^{-1}(\mathbf{M}_g) $. As $ \mathcal{Q} $ is a connected manifold, a smooth interpolation curve (e.g., a geodesic or polynomial spline) connecting $ \mathbf{q}_0 $ and $ \mathbf{q}_g $ can be constructed within it. The image of this curve under $ f $ yields the desired morphological evolution path. Uniqueness is guaranteed by the injectivity of $ f^{-1} $.

	\textbf{Predicting Motion Commands from Noisy Observations.} By maximizing the likelihood function estimation under noisy observations, motion commands are predicted from input observations corrupted by noise $ \epsilon $, subject to the morphological control constraint that the position error satisfies $ \|\mathbf{q}_g - \mathbf{q}^*\| < e $:

	\begin{equation}
		\underset{\Theta}{\max} \mathbb{E}_{(M_0, M_g, q^*) \sim \mathcal{D}} [\text{log} \, p_{\Theta}(\mathbf{q}^*| \mathbf{z}_0, \mathbf{z}_g)].
	\end{equation}

	Maximizing the log-likelihood is equivalent to minimizing the Kullback-Leibler divergence between the predicted distribution and the true distribution \cite{seghouane2007aic, hua2017matrix,van2014renyi}. Under the assumption of Gaussian observation noise and a Gaussian predictive distribution \cite{goodman1963statistical, novey2009complex}, this optimization problem reduces to a regularized mean squared error minimization problem:

	\begin{equation}
		\min_{\Theta} \mathbb{E}_{(M_0, M_g, q^*) \sim \mathcal{D}} \left[ \| \mathbf{q}^* - f_{\Theta}(\mathbf{z}_0, \mathbf{z}_g) \|^2 + \lambda \cdot \mathcal{R}(\Theta, \sigma_{\epsilon}) \right].
	\end{equation}

	The mean squared error term $ \| \mathbf{q}^* - f_{\Theta}(\mathbf{z}_0, \mathbf{z}_g) \|^2 $ measures the discrepancy between the predicted motion command $ f_{\Theta}(\mathbf{z}_0, \mathbf{z}_g) $ and the ground truth motion command $ \mathbf{q}^* $. The regularization term $ \lambda \cdot \mathcal{R}(\Theta, \sigma_{\epsilon}) $ accounts for the uncertainty introduced by the variance $ \sigma_{\epsilon}^2 $ of the observation noise $ \epsilon $ under the Gaussian assumption.

\section{Morphology-Based Whole-Body Motion Control}

\subsection{Discretized Representation of Robot Morphology}

	The key to entity discretization lies in ensuring that the represented topological features maintain spatiotemporal consistency. As invariants of embodiment morphology and motion, these topological features manifest concretely in Cartesian space as the robot's joint motions constrained by such features. The representation of fixed topological features by MPs enables the MophoGuard to learn the motion characteristics of the robots.

	To this end, we propose an entity discretization representation method that couples a MPs with e-skin (e-skin). The large-area distributed e-skin can cover the entire robot body, with its units closely attached to the robot skeleton, thereby maintaining the consistency of relative positions among components. By coupling each unit with MPs and ensuring spatiotemporal consistency, the spatial position tracking of the MPs achieves analytical-level accuracy through forward solving of each unit's position. This provides a solid foundation for the collection of high-quality MorphoGuard datasets and for model training.

	We apply this discretization representation method to a dual-arm robotic platform. The platform is fully covered with e-skin units and is used for data collection and model evaluation.

\subsection{Model Architecture}

	MorphoGuard aims to predict the relative joint configuration vector $\Delta \mathbf{q}$ corresponding to the robot's transition from its current morphology $M_0$ to the target morphology $M_g$. The network takes as input the current morphology of the MPs $M_0$ and the difference between the current morphology and the target morphology $\Delta M = M_g - M_0$.

	We design MorphoGuard as an encoder-decoder neural network. To address the challenge of significant modality gap between configuration space and morphology, which limits model comprehension, the architecture follows a feature extraction, information fusion, and joint command decoding pipeline, comprising modules for current morphology encoding, target morphology encoding, information fusion, backbone decoding, and action output. An overview of the overall network architecture is shown in Fig. \ref{fig:2}; further details can be found in the model recommendation experiments.

\subsection{Dataset Collection}

\begin{figure}[htbp]
	\centering
	\includegraphics[width= 0.8 \textwidth]{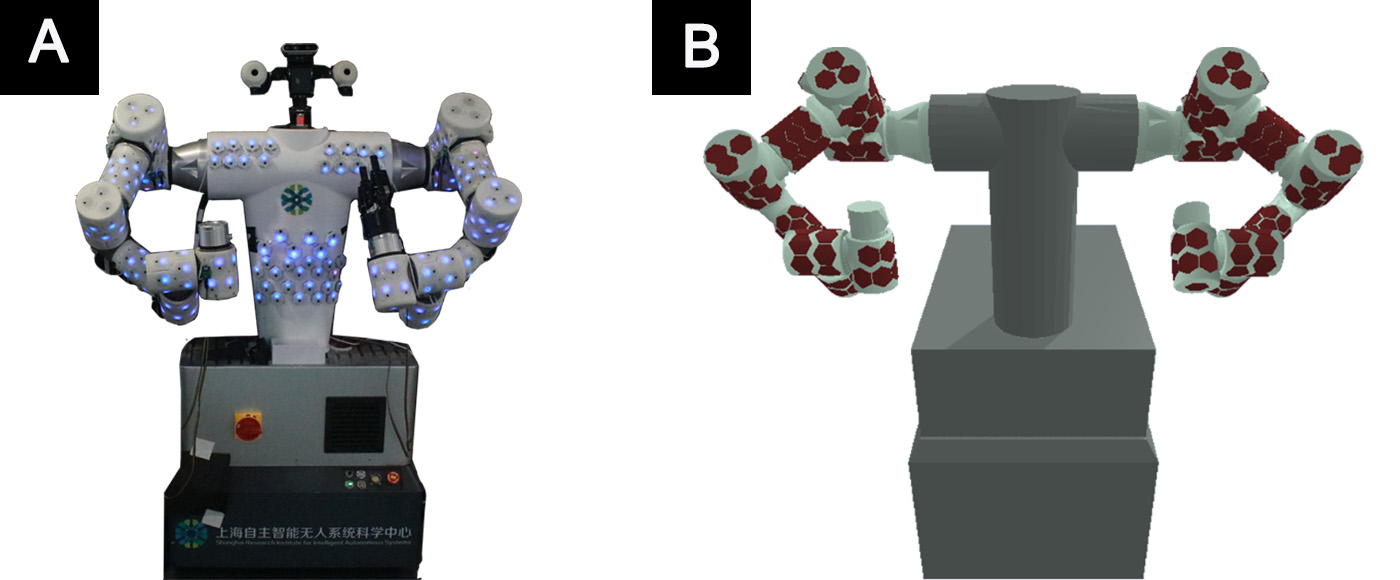} 
	\caption{Simulation and physical platform used for data collection.}
	\label{fig:3}
	\end{figure}

	To obtain sufficient robotic limb motion data for training MorphoGuard, we constructed simulated and physical robotic systems with consistent component configurations (Fig. \ref{fig:3}). Using a hybrid data collection approach, we gathered pairs of joint angles and material point system spatial positions (corresponding to e skin unit locations) within the robot's operational space.

	We calibrated the position of each e skin unit relative to its parent joint. Accordingly, the morphology $M$ of the robot in configuration space $\mathbf{q}$ can be expressed as $M = T(\mathbf{q}) P_e$, where $T(\mathbf{q})$ is the homogeneous transformation matrix of the e skin system with respect to the world coordinate system.

	To enable the robot to traverse its workspace and acquire joint-morphology pairs, we randomly generated a large number of discrete spatial target points within the workspace, with a spacing of 2 $mm$ between consecutive points. After organizing these target points into a sequence, the robot solved the inverse kinematics using the Jacobian inverse method, thereby traversing the target points sequentially. We collected a total of 1.3 million joint-morphology pairs.

\subsection{Datasets building}
	Based on the collected bodily motion data, we constructed a relative displacement dataset. Using a sampling method, we extracted two data pairs representing the current bodily state and the target motion state, respectively. The morphology and joint angles of the target state were each subtracted from those of the current state to obtain a morphology motion vector and a joint motion vector, with the latter serving as the label for joint control prediction. The sampling interval between the two data pairs was determined using a mixture of Gaussian distributions, with intervals ranging from 10 to 150 steps (as shown in Fig. 2a). This mixed sampling method introduces greater motion diversity, thereby enhancing the model's generalization performance.

	After dataset construction, given that both datasets were exceptionally large in scale, we allocated 98 $\%$ to the training set, 1$\%$  to the validation set, and the remaining 1$\%$  to the test set. The sizes of the datasets are shown in Fig. 2. Fig. 2c presents the distribution of motion features across the training, validation, and test sets for each dataset. Statistical results indicate that the three sets maintain a high degree of consistency, while also ensuring that the test data remain completely unseen during training.

\subsection{Training Objective}

	The training objective $ L_{\text{total}} $ is formulated as a dual-objective loss function, comprising motion feature learning and morphology noise estimation.

	Motion feature loss $ L_m $: This loss measures the error between the model prediction and the ground-truth motion command:
	\begin{equation}
	L_m = \|\mathbf{q}^* - f_{\Theta}(\mathbf{z}_0, \mathbf{z}_g)\|^2.
	\end{equation}

	Morphology noise loss $ L_g $: Under the Gaussian noise assumption, this loss measures the error between the predicted noise and the ground-truth morphology noise:
	\begin{equation}
	L_g = \lambda \cdot \mathcal{R}(\Theta, \sigma_{\epsilon}).
	\end{equation}

	The overall training objective is the weighted sum of the two losses, with weights $\lambda_1$ and $\lambda_2$ respectively:
	\begin{equation}
	L_{\text{total}} = \lambda_1 L_m + \lambda_2 L_g.
	\end{equation}

\section{Model Recommendation Experiments}

	This experiment aims to identify the key factors influencing MorphoGuard performance by establishing and analyzing the performance boundaries of baseline models, thereby proposing a recommended model. The experiment adopts a progressive complexity and modular ablation approach to design a series of baseline models. By observing the magnitude of performance gains, we quantitatively evaluate the contribution of different architectural designs to model performance.

\subsection{Comparative Study of Backbones}

	To validate the alignment between the inductive biases of different network architectures and the underlying data structure of the WBC task, we first conducted an inductive bias adaptation stress test. Following a unified network framework, we selected four representative backbone networks—MLP, CNN, Transformer, and GNN—as feature extractors (as shown in Table \ref{tab:architecture_comparison}). While identifying the best-performing backbone based on performance ceiling and convergence efficiency, we place greater emphasis on interpreting its success from the perspective of inductive bias, thereby providing theoretical guidance for subsequent architecture design tailored to this task.


	\begin{table}[htbp]
	\centering
	\small
	\caption{Architecture Comparison - Core Metrics Focus}
	\label{tab:architecture_comparison}
	\begin{tabular}{p{2.1cm} p{3.5cm} p{4cm} p{1.1cm} p{1.1cm}}
	\toprule
	\textbf{Architecture} & \textbf{Core Inductive Bias} & \textbf{Configuration} & \textbf{Params} & \textbf{FLOPs} \\
	\midrule
	MLP & Global Feature Combination & 4-layer MLP, 512-512-256-128 & 1.5M & 12M \\
	CNN & Locality + Translation Invariance & ResNet-18 & 11M & 1.8G \\
	Transformer & Long-range Dependencies & 6 layers, 4 heads, 512 dim & 65M & 3.2G \\
	GNN & Graph Structure Propagation & 4-layer GAT, 128 dim & 0.8M & 8M \\
	\bottomrule
	\end{tabular}
	\end{table}

	Experimental results demonstrate that MLP Robot Policy achieves the lowest validation loss (0.0396) among all five backbone networks. Although its training loss (0.0066) is slightly higher than Transformer (0.0047) and GNN (0.0051), the validation loss significantly outperforms other methods. The validation loss standard deviation of MLP (0.0169) is relatively larger, indicating slightly lower stability across training epochs, yet this variability does not compromise its overall performance advantage. In comparison, CNN, GNN, and Transformer all exhibit validation losses above 0.046, with negative relative improvements ranging from -16.98$\%$ to -18.37$\%$ compared to the MLP baseline, with only CNN showing marginal significance ($p=0.0864$). Considering the minimization of validation loss, architectural simplicity, and computational efficiency, selecting MLP Robot Policy as the backbone network is well-justified.

	\begin{figure}[htbp]
	\centering
	\includegraphics[width= \textwidth]{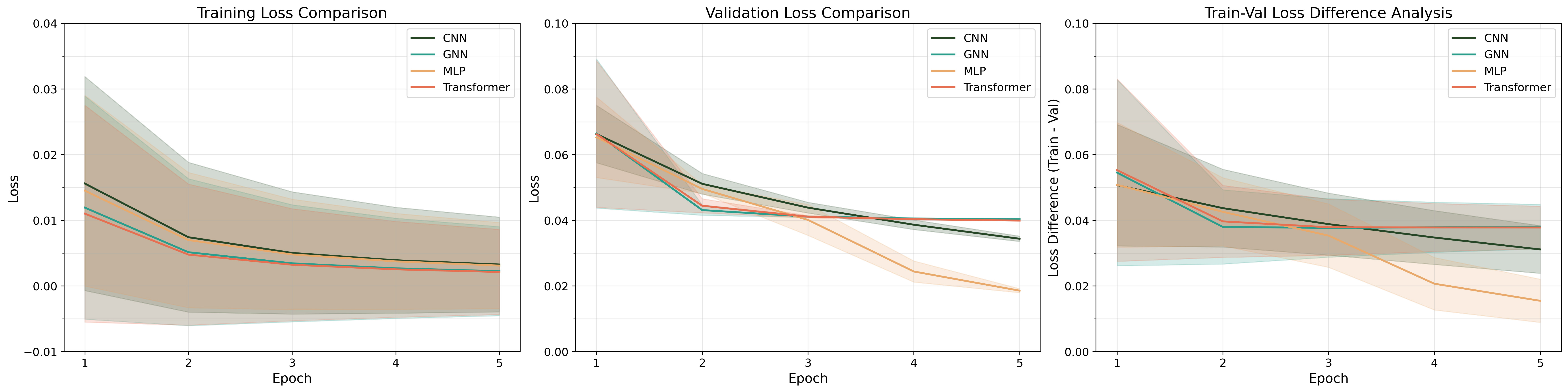} 
	\caption{Training and validation loss curves for different backbone architectures. The MLP backbone demonstrates the lowest validation loss, indicating superior generalization performance compared to CNN, Transformer, and GNN backbones.}
	\label{fig:4}
	\end{figure}

\subsection{Model Capacity Scaling Study}

	Given the importance of model computational efficiency in robotic motion control, we further designed an experiment to evaluate the relationship between model parameter count and learning performance. This experiment quantifies the trade-off between model performance and computational cost, thereby enabling the identification of optimal architectural configurations in resource-constrained real-world applications. We systematically scale the depth (number of layers) and width (hidden dimension) of the MLP backbone to construct a series of model variants with progressively increasing parameter counts, ranging from small models of 5M parameters to large models exceeding 200M parameters. To enable the model to better accommodate deep architectures, we introduce residual modules. The performance of the models evaluated in the tests is shown in Table \ref{tab:mlp_scaling}. During the experiment, we meticulously record for each variant the core task metrics on the validation set, training convergence speed, per-step iteration time, inference latency, and peak GPU memory usage.


	\begin{table}[htbp]
	\centering
	\caption{MLP Scaling Experiment: Model Architecture and Computational Efficiency}
	\label{tab:mlp_scaling}
	\begin{tabular}{lcccc}
	\hline
	\textbf{Model Name} & \textbf{Params (M)} & \textbf{Backbone Config} & \textbf{FLOPs (G)} & \textbf{Inference Time (ms)} \\
	& & \textbf{(Layers $\times$ Width)} & & \textbf{(batch=64)} \\
	\hline
	small\_5m & 5.9 & 8 $\times$ 512 & 0.75 & 0.8--1.2 \\
	medium\_10m & 10.2 & 8 $\times$ 768 & 1.95 & 1.2--1.5 \\
	medium\_25m & 25.9 & 10 $\times$ 1024 & 3.95 & 4.5--6.5 \\
	large\_50m & 51.9 & 22 $\times$ 1024 & 6.64 & 7.5--10.5 \\
	large\_100m & 115.4 & 22 $\times$ 1536 & 15.05 & 18--25 \\
	xlarge\_200m & 223.1 & 24 $\times$ 2048 & 30.59 & 35--50 \\
	\hline
	\end{tabular}
	\footnotesize
	\textit{Note:} Inference times are measured on NVIDIA RTX 4090 with batch size 64. FLOPs are calculated for forward pass (inference).
	\end{table}

	Through systematic model scaling analysis (Table \ref{tab:scale_comparison_results}), we identify the 25M parameter configuration as the optimal model size for this task. This configuration achieves a 60.27$\%$ relative improvement in validation loss ($p=0.0204, p=0.0204$) compared to the 5M baseline, representing the point where performance gains begin to plateau. Larger models (50M-200M) yield diminishing returns, with the 200M configuration even showing slight performance degradation. The 25M configuration thus offers the best balance between performance improvement and computational efficiency.

	\begin{table}[htbp]
	\centering
	\caption{Model Scale Comparison Results}
	\label{tab:scale_comparison_results}
	\begin{tabular}{l c c c c c}
	\toprule
	\textbf{Model Size} & \textbf{Val Loss} & \textbf{Val Std} & \textbf{Rel Improvement (\%)} & \textbf{p-value} & \textbf{Sig} \\
	\midrule
	Scale Small 5M & 0.0292 & 0.0229 & 0.00 & nan & ns \\
	Scale Medium 10M & 0.0178 & 0.0208 & 39.02 & 0.0204 & * \\
	\bfseries Scale Medium 25M & \bfseries0.0116 & \bfseries0.0151 & \bfseries60.27 & \bfseries0.0204 & \bfseries* \\
	Scale Large 50M & 0.0096 & 0.0127 & 67.18 & 0.0257 & * \\
	Scale Large 100M & 0.0095 & 0.0120 & 67.50 & 0.0291 & * \\
	Scale Xlarge 200M & 0.0104 & 0.0133 & 64.36 & 0.0234 & * \\
	\bottomrule
	\end{tabular}
	\vspace{2mm}\\
	{\footnotesize \textbf{Note:} Val Loss = Validation Loss, Val Std = Standard Deviation, Rel Improvement = Relative Improvement compared to Scale Small 5M baseline. Significance levels: * p $<$ 0.05, ns = not significant. \textbf{Bold} row indicates the optimal configuration.}
	\end{table}

\subsection{Progressive Feature Fusion Analysis}

	After identifying an efficient model backbone (derived from the backbone experiment) and optimal computational configuration (obtained from the scaling experiment), the model's performance on the existing feature set has reached saturation. To further push the performance ceiling, we shift focus to the input side and investigate the impact on model performance ranging from feature concatenation to deep fusion.

	To this end, we design a progressive fusion evaluation scheme to examine the effect of moving from simple concatenation to deep interaction. First, as a baseline, we adopt a \textbf{feature concatenation (concat)} approach that directly concatenates the state vector and the motion vector as network input. Second, we introduce feature \textbf{crossinteraction} by computing the outer product of the two vectors to explicitly capture their nonlinear interactions. Finally, we employ a feature \textbf{modulation (Parallel Branch)}	 strategy based on the FiLM mechanism, where the motion vector serves as a condition to dynamically modulate the processing of the state features, enabling a deeper level of fusion.

	Table \ref{tab:fusion_results} presents a comparative evaluation of five fusion methods. Among all evaluated strategies, Additive fusion achieves the lowest validation loss (0.0159) with a standard deviation of 0.0211. Alternative methods including Concat (0.0232, -45.81$\%$), Input Concat (0.0171, -7.86$\%$), Parallel Branch (0.0321, -102.05$\%$), and Adaptive Norm (0.0178, -11.90$\%$) all underperform the additive baseline. Notably, while Parallel Branch shows statistical significance ($p=0.0407$), its performance is substantially worse, indicating that increased complexity does not yield improvements. The remaining methods show no significant differences from Additive fusion ($p>0.05$), yet consistently exhibit higher validation losses. These results demonstrate that the simple element-wise addition of features provides the most effective fusion mechanism for this task. Consequently, we adopt Additive fusion as the baseline for all subsequent experiments.

	\begin{table}[htbp]
	\centering
	\small
	\caption{Fusion Method Comparison Results}
	\label{tab:fusion_results}
	\begin{tabular}{l c c c c c}
	\toprule
	\textbf{Fusion Method} & \textbf{Val Loss} & \textbf{Val Std} & \textbf{Rel Improvement (\%)} & \textbf{p-value} & \textbf{Sig} \\
	\midrule
	\bfseries Additive & \bfseries 0.0159 & \bfseries 0.0211 & \bfseries 0.00 & \bfseries --- & \bfseries --- \\
	Concat & 0.0232 & 0.0248 & -45.81 & 0.1112 & \\
	Input Concat & 0.0171 & 0.0230 & -7.86 & 0.2473 & \\
	Parallel Branch & 0.0321 & 0.0218 & -102.05 & 0.0407 & $*$ \\
	Adaptive Norm & 0.0178 & 0.0229 & -11.90 & 0.1015 & \\
	\bottomrule
	\end{tabular}
	\vspace{2mm}\\
	{\footnotesize \textbf{Note:} Baseline method (Additive) shown in \textbf{bold}. $*$ indicates statistical significance at p $<$ 0.05.}
	\end{table}

\section{Evaluation}
\label{sec:result}

	In this experiment, we introduce a multi object manipulation task designed to evaluate the performance of MorphoGuard in whole body interaction management. The task requires the robot to use any part of its body to place one or more objects at target positions within a permissible contact error tolerance.

\begin{figure}[htbp]
	\centering
	\includegraphics[width= 0.8 \textwidth]{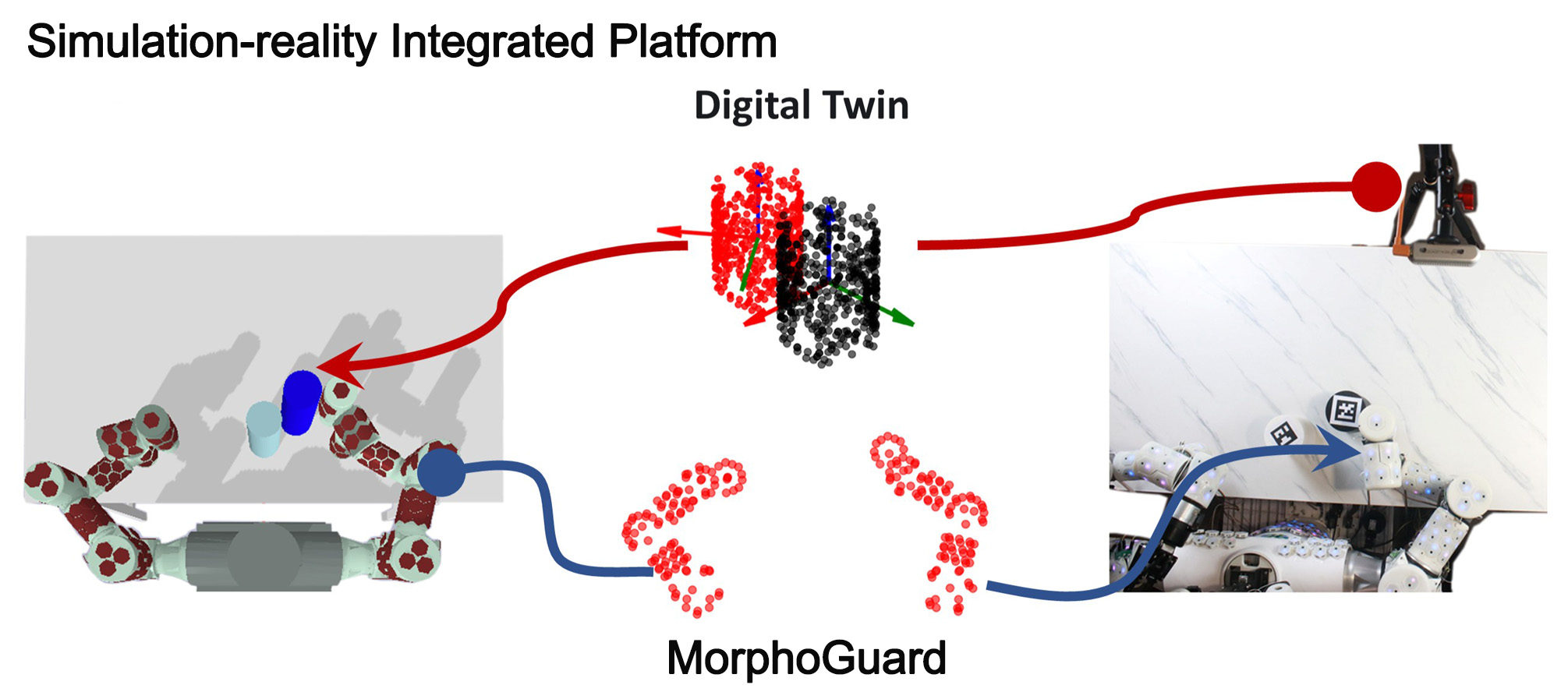} 
	\caption{}
	\label{fig:5}
	\end{figure}

	We construct a simulation-reality integrated test platform to support the execution of this task (Fig. \ref{fig:5}). The platform generates the spatial states of MPs in a simulation environment, which are then fed into MorphoGuard to produce joint commands that drive the robot to complete the multi object manipulation task (figure). Details regarding platform construction are provided in Appendix 3. We evaluate the model's whole body interaction performance using the metric of contact point control accuracy.

\begin{figure}[htbp]
	\centering
	\includegraphics[width= \textwidth]{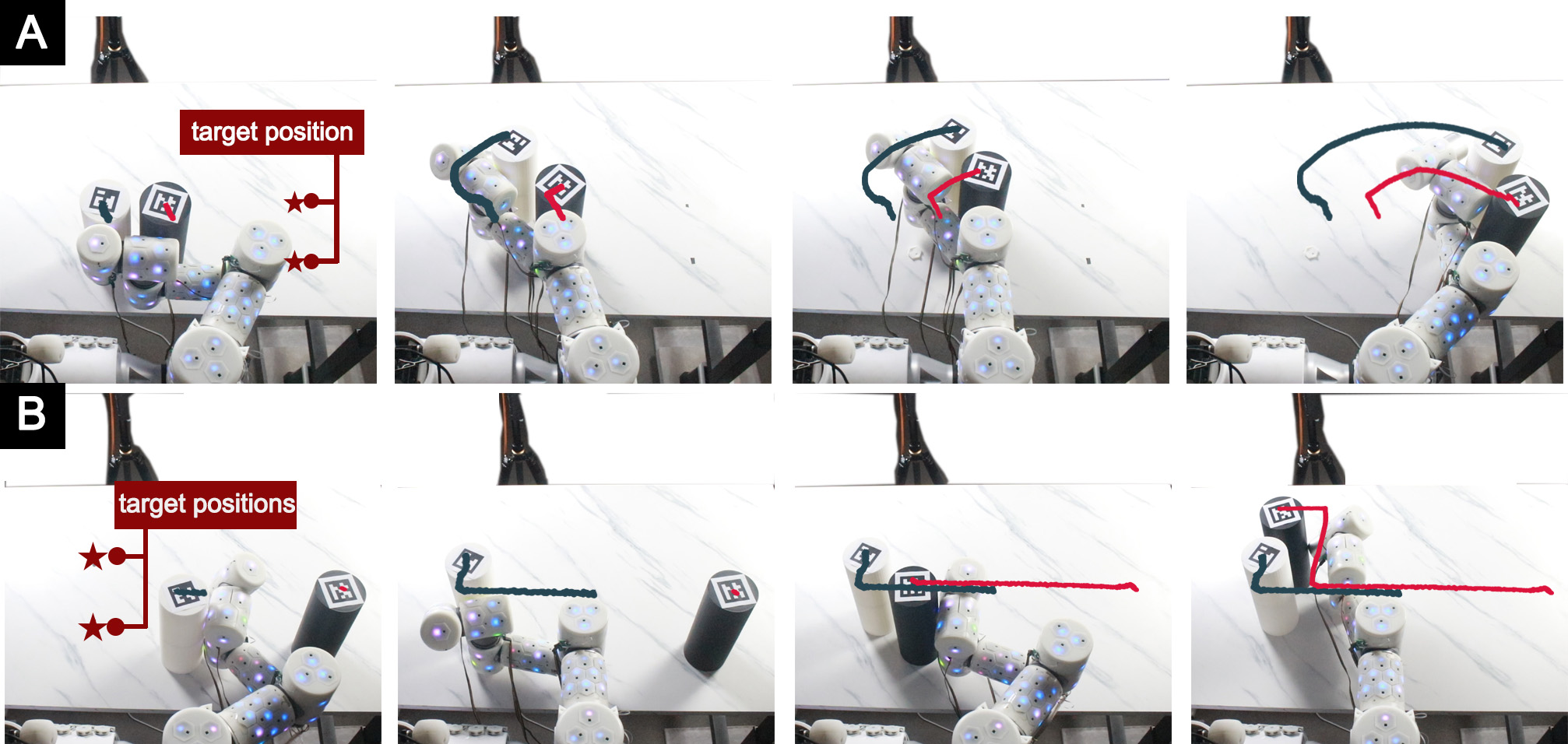} 
	\caption{Test procedure and object trajectories for the sequential manipulation task. (A)sequential manipulation task. (B)simultaneous manipulation task.}
	\label{fig:6}
	\end{figure}

	We conducted two tests. In the first test, the robot was required to manipulate two objects sequentially to their target positions. The first object interfered with the motion of the second, requiring the robot to navigate around the first object to successfully deliver the second. The test procedure and object trajectories are shown in Fig. \ref{fig:6}A. The robot successfully manipulated both objects to their target positions without any collisions (see Supplementary Video 1). As illustrated in Fig. \ref{fig:7}A, the contact point error between the robot and objects in the physical environment relative to the virtual scene ranged from 0 to 2 cm, indicating that MorphoGuard exhibits robust contact point management capability.

	In the second test, the robot was required to manipulate two objects simultaneously to their target positions. Fig. \ref{fig:6}B depicts the task execution process and the corresponding object trajectories. Both objects reached their target positions concurrently under the robot's manipulation. The contact position error between the robot and objects in the physical environment relative to the simulated environment ranged from 0 to 1.2 cm, as shown in Fig. \ref{fig:7}B. These results demonstrate that MorphoGuard, based on material point representation, can accomplish multi-contact management tasks by predicting joint commands that track the spatial states of the material point system (see Supplementary Video 2).

	\begin{figure}[htbp]
	\centering
	\includegraphics[width= 0.8 \textwidth]{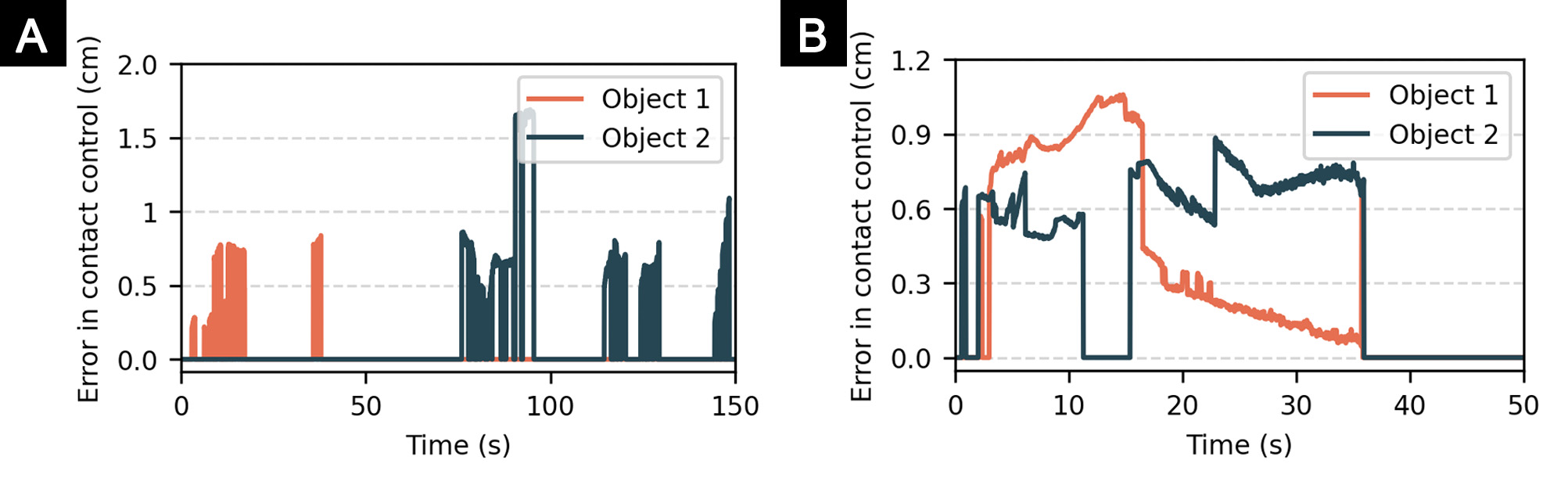} 
	\caption{Test results for the multi-object manipulation tasks. (A) Contact point error for the sequential manipulation task. (B) Contact point error for the simultaneous manipulation task.}
	\label{fig:7}
	\end{figure}


\section{Conclusion}
\label{sec:conclusion}

	In this work, we introduce MorphoGuard, a morphology-based whole-body motion control method that leverages material point representation and neural network modeling to achieve robust and efficient multi-contact control of complex robotic systems. Through systematic model recommendation experiments, we identify the optimal architecture and configuration for MorphoGuard, demonstrating its superior performance in multi-object manipulation tasks. The results highlight the potential of morphology-based approaches in advancing whole-body interactive motion control for robotics.

\bibliographystyle{unsrt}  
\bibliography{references}  

\end{document}